\documentclass[11pt,a4paper]{article}

\usepackage[english]{babel}

\usepackage[T1]{fontenc}
\usepackage[math]{iwona} 

\usepackage{amsmath}

\usepackage{graphicx}

\usepackage{placeins}

\usepackage{natbib} 
\usepackage{IEEEtrantools}

\usepackage{tikz} \usetikzlibrary{shapes}

\usepackage{hyperref} 

\usepackage{calc}

\usepackage[paper=a4paper,portrait, tmargin=30mm, bmargin=30mm, rmargin=40mm, lmargin=25mm]{geometry} 

\usepackage{tabls}

\newlength\micolumna
\title{ISP pricing and Platform pricing interaction under net neutrality}

\author{Luis Guijarro, Vicent Pla, José Ramón Vidal\\
Universitat Politècnica de València}

\date{}

\begin{document}

\thispagestyle{empty}

\begin{tabular*}{\textwidth}[t]{|p{\micolumna}|p{\micolumna}|}
	\hline
	\includegraphics[width=\micolumna,trim=2cm 17cm 2cm 2cm,clip]{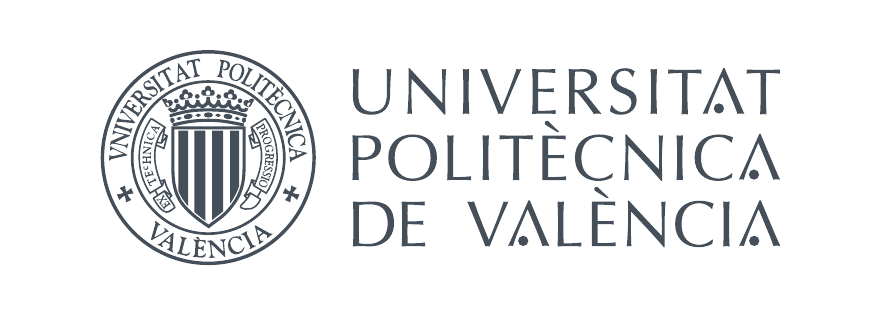}& 
	\raisebox{1.6cm}{\parbox{\micolumna}{Departamento de Comunicaciones}}\\
	\hline
	\multicolumn{2}{|c|}{}\\
	[0.20\textheight]
	\multicolumn{2}{|p{\textwidth-\tabcolsep*2}|}
	{\huge \begin{center}
ISP pricing and Platform pricing interaction under net neutrality
\end{center}}\\
	\multicolumn{2}{|c|}{\huge (Document NETECON134F4)}\\
	[0.20\textheight]

	\hline
	Authors: Luis~Guijarro,  & Outreach: Public\\ 
	\cline{2-2}
	Vicent Pla, José Ramón Vidal & Date: December 11, 2023 \\ 
	\cline{2-2}
	 & Version: b\\
	\hline
\end{tabular*}

\clearpage

\setcounter{page}{1}

\maketitle

\begin{abstract}
We analyze the effects of enforcing vs. exempting access ISP from net neutrality regulations when platforms are present and operate two-sided pricing in their business models. This study is conducted in a scenario where users and Content Providers (CPs) have access to the internet by means of their serving ISPs and to a platform that intermediates and matches users and CPs, among other service offerings.
Our hypothesis is that platform two-sided pricing interacts in a relevant manner with the access ISP, which may be allowed (an hypothetical non-neutrality scenario) or not (the current neutrality regulation status) to apply two-sided pricing on its service business model.
We preliminarily conclude that the platforms are extracting surplus from the CPs under the current net neutrality regime for the ISP, and that the platforms would not be able to do so under the counter-factual situation where the ISPs could apply two-sided prices.
\end{abstract}

\section{Introduction}

Net neutrality has been debated intensively since it first was advocated two decades ago. And it has been regulated worldwide, prominently in the US and in the EU. There are multiple approaches to the net neutrality concept. We focus on the one which prevents a two-sided pricing scheme to be applied by an access Internet Service Provider (ISP) in order for Content Providers (CPs) reaching the access ISP's subscribers from a different ISP~\citep{economides2012}.

It has been claimed by access ISPs that charging a side payment to the CPs, which is forbidden by net neutrality regulations, would contribute to the upgrade of the infrastructure needed to support the huge amount of traffic that flows from the CPs to the users. This rationale has recently been proposed under the concept of ``direct compensation'' or ``fair share''.

We do not aim to contribute to the general debate on net neutrality under the current facade~\citep{stocker2022}. Instead, we will focus on the fact that CPs (e.g., newspapers) not only need access ISP in order to reach users that subscribe to their services, but also need platforms (e.g., Google News) that intermediate and match CPs against users. These platforms do not abide to an equivalent net neutrality regulation, and therefore have been applying two-sided pricing mechanisms, which allows them to actively manage the cross-network effects operating in such business models.

Our focus is then to analyze the effects of enforcing vs. exempting access ISP from net neutrality regulations when platforms are present and operate two-sided pricing in their business models. Our hypothesis is that platform two-sided pricing interacts in a relevant manner with the access ISP, which may be allowed (an hypothetical non-neutrality scenario) or not (the current neutrality regulation status) to apply two-sided pricing on their service business model.

This work has been conducted during a research stay at the Weizenbaum Institut, Berlin, Germany, in July 2023, hosted by the Research Group \emph{Digital Economy, Internet, Ecosystems and Internet Policy} thanks to Dr.~Stocker's invitation. Financial support is acknowledged from Grant PID2021-123168NB-I00, funded by MCIN/AEI, Spain/10.13039/ 501100011033 and the European Union A way of making Europe/ERDF.

\section{Economic Model}\label{sec:model}

\begin{figure}[t]
\begin{center}
\includegraphics[width=0.6\textwidth]{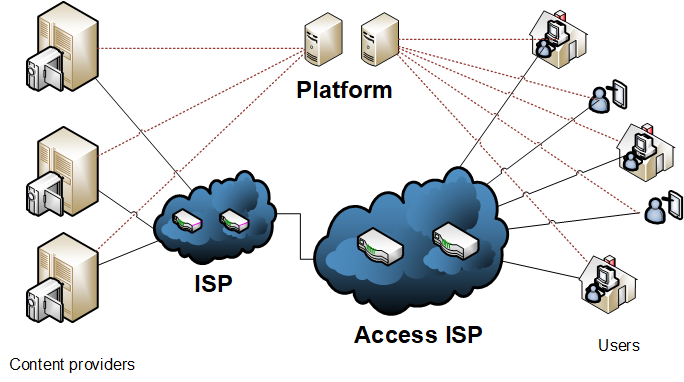}
\caption{Scenario}\label{fig:scenario}
\end{center}
\end{figure}

We model a scenario as depicted in Fig.~\ref{fig:scenario}, where users and CPs have access to the internet by means of their serving ISPs and to a platform that intermediates and matches users and CPs, among other service offering. Therefore, both ISPs and platform are necessary for the users to subscribe and use the services provided by the CPs. And both ISPs and platform create their respective stand alone value with additional (although typically more basic) services.

\subsection{Users' subscription}

We model a mass of $N_u$ non-atomic users that are potential subscribers of both the access ISP and the platform, so that if they do subscribe to both of them, they will be able to enjoy the service provided by the CPs. The users are homogeneous in all their characteristics except in the value of the outside option, which we assume is drawn independently from a uniform distribution over the unity interval $[0,1]$.

The users are charged a fee $\beta$ by the platform and a price $b'$ by the access ISP, per traffic unit that is downloaded from the CPs. If an average traffic $\omega$ is assumed for the traffic downloaded by one user from one CP, then we can set $b=b' \omega$ as the price per CP that the access ISP charges to a user.

The users derive a stand-alone value $r_u$ from the combined platform-access-ISP service. And they derive an additional value that is increasing in the number of CPs offering their contents through the platform and the access ISP. If the number of joined CPs is $n_c$, this additional value, assuming a linear dependence, is $\delta n_c$, so that $\delta$ models the intensity of the cross-group network effect that the CP side exerts on the users. 

Putting all the above modeling decisions together, the expression for the utility that a user derives if he/she subscribes to the combined service is:
\begin{equation}
u = r_u + \delta n_c - \beta - b n_c = r_u - \beta + (\delta - b) n_c
\end{equation}

Note that the above expression has similarities with the common modeling of the utility derived by the users of a platform when they are charged both a participation fee $\beta$ and a transaction fee $b$~\citep{rochet2006}. Nevertheless, in this work, as it will be detailed below, these fees are charged by different agents.

Finally, since the outside option $u_0$ of each user is uniformly distributed in the unity interval, the number of users that will subscribe to the combined service $n_u$ can be computed as:
\begin{align}\label{eq:users}
\frac{n_u}{N_u} & =\text{Prob} \{u_0 \leq u \} = \nonumber\\ 
& = \text{Prob}\{ u_0 \leq  r_u - \beta + (\delta - b) n_c \} \nonumber  \\
& = \begin{cases}
 0 & \text{if }   r_u - \beta + (\delta - b) n_c < 0\\
  r_u - \beta + (\delta - b) n_c & \text{if }  0  \leq r_u - \beta + (\delta - b) n_c \leq 1\\
 1 &  \text{if }  1  < r_u - \beta + (\delta - b) n_c
\end{cases}
\end{align}

\subsection{CPs' decisions}

We model a mass of $N_c$ non-atomic CPs that are willing to offer their contents to the users, which are reachable by means of the combined platform-ISPs service. The CP's business model is based on advertisement. The CPs are homogeneous in all their characteristics except in the benefit of the outside option, which we assume it is drawn independently from a uniform distribution over the unity interval $[0,1]$.

A CP is charged a fee $\alpha$ by the platform and a price $a'$ per traffic unit that is uploaded to its ISP. Above we assumed that an average traffic $\omega$ is downloaded from a CP to a user, so that we can denote by $a=a' \omega$ the price per user that is charged to the CP. If the access ISP is allowed to apply a two-sided pricing mechanism, this will add an additional fee $c=c' \omega$ per user that the access ISP will charge to the CP.

The CP derives a stand-alone benefit $r_c$ from the combined platform-ISPs service. And it gets an advertising revenue $\gamma$ per user. Since the number of subscribers is $n_u$, this additional revenue is $\gamma n_u$, so that we can interpret $\gamma$ as the intensity of the cross-group network effects that the user side exerts on the CPs. 

Putting all the above modeling decisions together, the expression for the profit that a CP obtains if joins to the combined service is:
\begin{equation}
\Pi_c = r_c + \gamma n_u - \alpha - (a+c) n_u = r_c - \alpha + (\gamma - c - a) n_u
\end{equation}

Again, note that the above expression has similarities with a setting where the CPs are charged both a participation fee $\alpha$ and a transaction fee $a+c$.

Finally, since the outside option $\Pi_0$ of each CP is uniformly distributed in the unity interval, the number of CPs that will join to the combined service $n_c$ can be computed as:
\begin{align}\label{eq:CPs}
\frac{n_c}{N_c} & =\text{Prob} \{\Pi_0 \leq \Pi_c \} = \nonumber\\ 
& = \text{Prob}\{ \Pi_0 \leq  r_c - \alpha + (\gamma - c - a) n_u \} \nonumber  \\
& = \begin{cases}
 0 & \text{if }   r_c - \alpha + (\gamma - c - a) n_u < 0\\
  r_c - \alpha + (\gamma - c - a) n_u & \text{if }  0  \leq r_c - \alpha + (\gamma - c - a) n_u \leq 1\\
 1 &  \text{if }  1  < r_c - \alpha + (\gamma - c - a) n_u
\end{cases}
\end{align}

\subsection{Platform's decisions}

The platform charges a fee $\beta$ to each subscriber and a fee $\alpha$ to each joined CP, so that it gets a revenue equal to
\begin{equation}
\Pi_p = \alpha n_c + \beta n_u.
\end{equation} 
We neglect the variable costs incurred by the platform, so that the platform will set the two-sided price $\{\alpha,\beta\}$ in order to maximize $\Pi_p$.

If the platform is absent, then fees $\beta$ and $\alpha$ are set to zero.

\subsection{ISP's decisions}

We assume that the ISP providing acces to the CPs is not an active agent in our model, so that we take $a$ as a parameter, and therefore its benefits are set to $a n_c$.

As regards the access ISP, if a non neutral two-sided pricing is allowed, its revenue will be given by:
\begin{equation}
\Pi_u = (b n_c) n_u + (c n_u) n_c = (b+c) n_u n_c,
\end{equation}
and the access ISP will set the pair $\{ b,c \}$ in order to maximize $\Pi_u$.

If the access ISP is instead under net neutrality regulation, which enforces $c=0$, the access ISP will only set $b$ in order to maximize $\Pi_u = b n_u n_c$.

\section{Analysis}\label{sec:analysis}

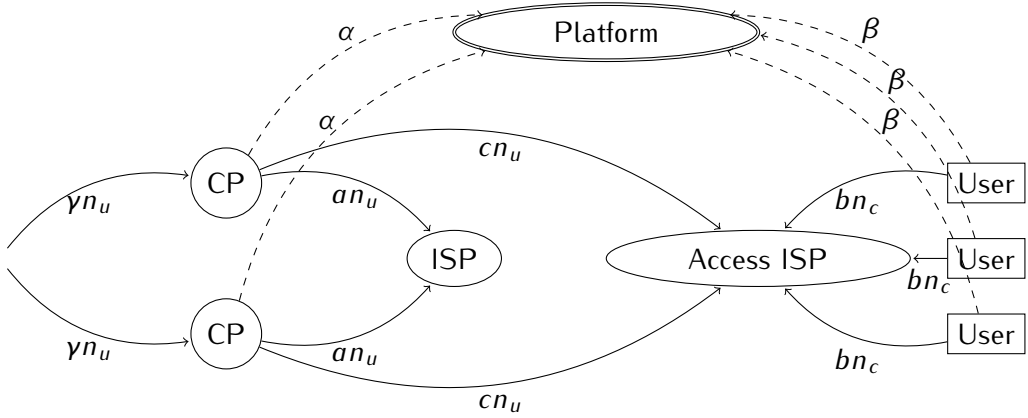
\begin{figure}[!t]
	\begin{center}
	\begin{tikzpicture}[node distance=3.0cm,shorten >=1pt] 
		\node[ellipse,draw,double,align=center, minimum width=4cm](platform){Platform};	
		\node[ellipse,draw,align=center](accessisp)[below of=platform, xshift=2cm, minimum width=4cm]{Access ISP};	
		\node[ellipse,draw,align=center](otherisp)[below of=platform, xshift=-2cm]{ISP};	
		\node[circle,draw,align=center](CP2)[left of=otherisp,yshift=-1cm]{CP};						\node[circle,draw,align=center](CP1)[left of=otherisp,yshift=1cm]{CP};
		\node(adv)[left of=CP1,yshift=-1cm]{};
		\node[rectangle,draw,align=center](user3)[right of=accessisp,yshift=-1cm]{User};
		\node[rectangle,draw,align=center](user2)[right of=accessisp,yshift=0cm]{User};
		\node[rectangle,draw,align=center](user1)[right of=accessisp,yshift=1cm]{User};
		\path[->]	(CP1) edge [bend left] node[below]{$a n_u$} (otherisp)
					(CP2) edge [bend right] node[below]{$a n_u$} (otherisp)
					(CP1) edge [bend left] node[below]{$c n_u$} (accessisp)
					(CP2) edge [bend right] node[below]{$c n_u$} (accessisp)
					(adv) edge [bend left] node[below]{$\gamma n_u$} (CP1)
					(adv) edge [bend right] node[below]{$\gamma n_u$} (CP2)
					(user1) edge [bend right] node[below]{$b n_c$} (accessisp)
					(user2) edge node[below]{$b n_c$} (accessisp)
					(user3) edge [bend left] node[below]{$b n_c$} (accessisp)
					(CP1) edge [bend left,dashed] node[above]{$\alpha$} (platform)
					(CP2) edge [bend left,dashed] node[above]{$\alpha$} (platform)
					(user1) edge [bend right,dashed] node[above]{$\beta$} (platform)
					(user2) edge [bend right,dashed] node[above]{$\beta$} (platform)
					(user3) edge [bend right,dashed] node[above]{$\beta$} (platform);
	\end{tikzpicture}
	\end{center}
	\caption{Platform and ISPs payment flow model.}\label{fig:payment}
\end{figure}

Fig.~\ref{fig:payment} summarizes the payments flow described above.

We assume the following sequence of decisions:
\begin{enumerate}
\item The access ISP sets $\{ c, b \}$
\item The platform sets $\{ \alpha, \beta \}$
\item The users and the CPs decide whether to subscribe/join or not.
\end{enumerate}

We therefore assume that the pricing decision by the access ISP is taken before the corresponding decision by the platform, under the assumption that the ISP's price decision is usually taken on a longer time frame that the pricing decision by a platform.

We do not restrict neither the platform nor the access ISP to set positive prices, so that they can set negative prices in one the side that creates stronger cross-network effects.

Furthermore, once prices are set by the access ISP and the platform, the subscription decision by users and CPs are modeled under the assumption of a fulfilled-expectations equilibrium, where agents (users or CPs) from one side form the same expectations on the participation of the agents of the other side and these expectations turn out to be correct. That is, the number of subscribers $n_u$ and joined CPs $n_c$ will be the solution $\{ n_u, n_c \}$ to the system of the two equations~\eqref{eq:users} and~\eqref{eq:CPs}~\citep[p.83]{belleflamme2021}.

To sum up: the platform sets  $\{ \alpha, \beta \}$ anticipating $\{ n_u, n_c \}$, and the access ISP sets $\{ c, b \}$ anticipating the platform decision and the resulting $\{ n_u, n_c \}$.

\section{Results}\label{sec:results}

\begin{figure}[t]
\begin{center}
\includegraphics[width=0.6\textwidth]{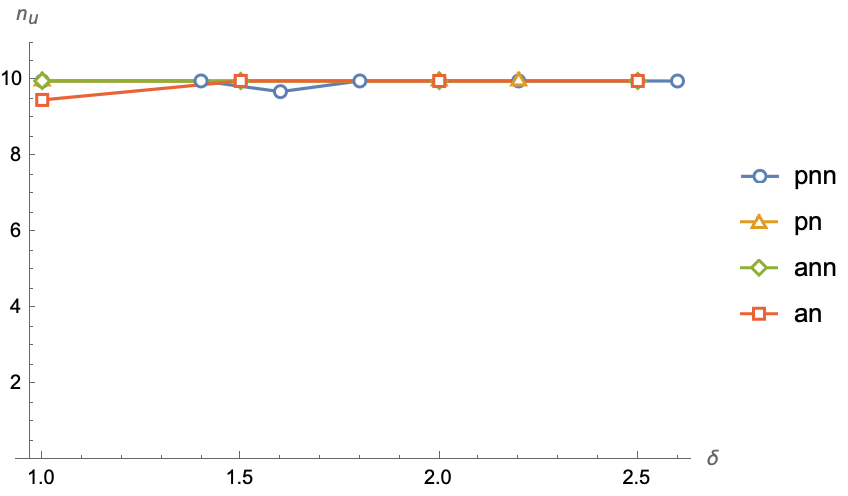}
\caption{Number of subscribers as a function of $\delta$}\label{fig:nudeltaplot}
\end{center}
\end{figure}

\begin{figure}[t]
\begin{center}
\includegraphics[width=0.6\textwidth]{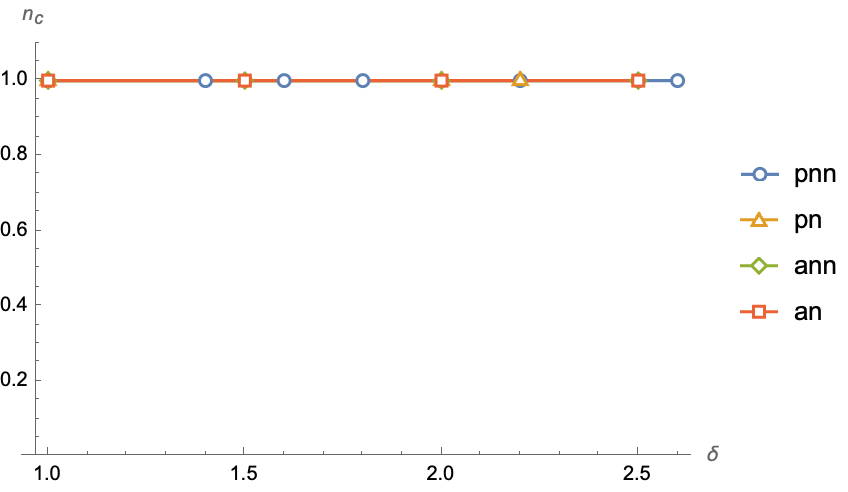}
\caption{Number of joined CPs as a function of $\delta$}\label{fig:ncdeltaplot}
\end{center}
\end{figure}

\begin{figure}[t]
\begin{center}
\includegraphics[width=0.6\textwidth]{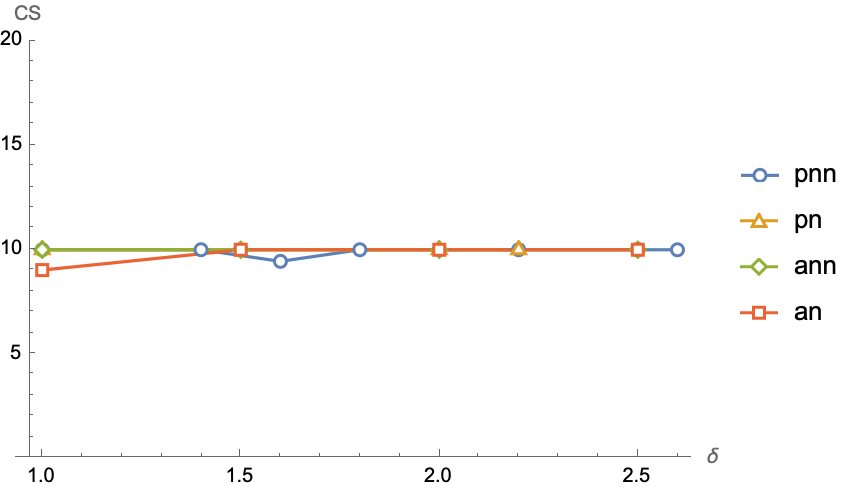}
\caption{Consumers' surplus as a function of $\delta$}\label{fig:CSdeltaplot}
\end{center}
\end{figure}

\begin{figure}[t]
\begin{center}
\includegraphics[width=0.6\textwidth]{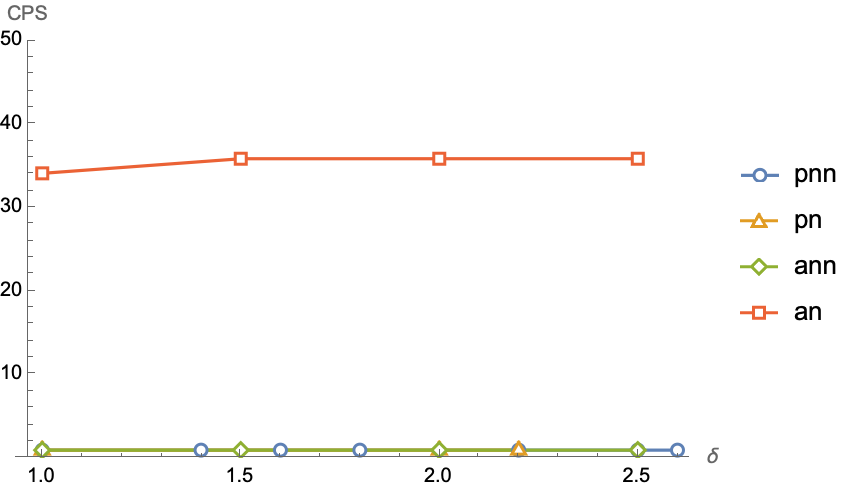}
\caption{CPs surplus as a function of $\delta$}\label{fig:CPSdeltaplot}
\end{center}
\end{figure}

\begin{figure}[t]
\begin{center}
\includegraphics[width=0.6\textwidth]{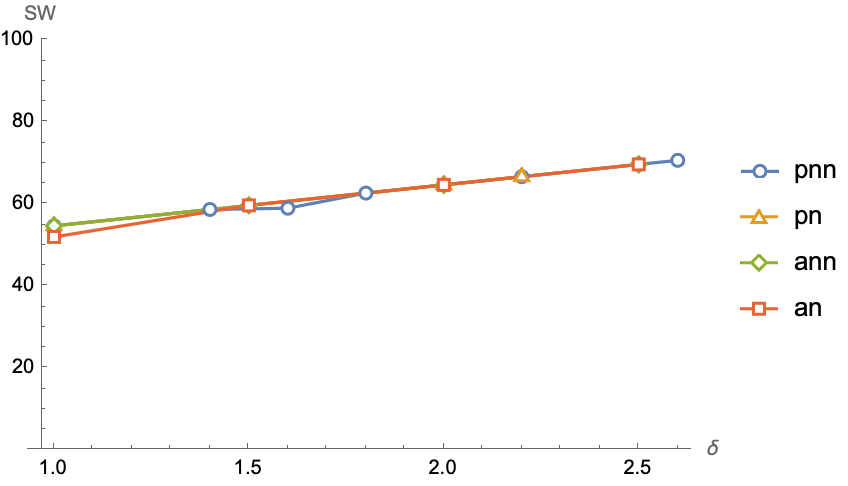}
\caption{Social welfare as a function of $\delta$}\label{fig:SWdeltaplot}
\end{center}
\end{figure}

\begin{figure}[thb]
\begin{center}
\includegraphics[width=0.6\textwidth]{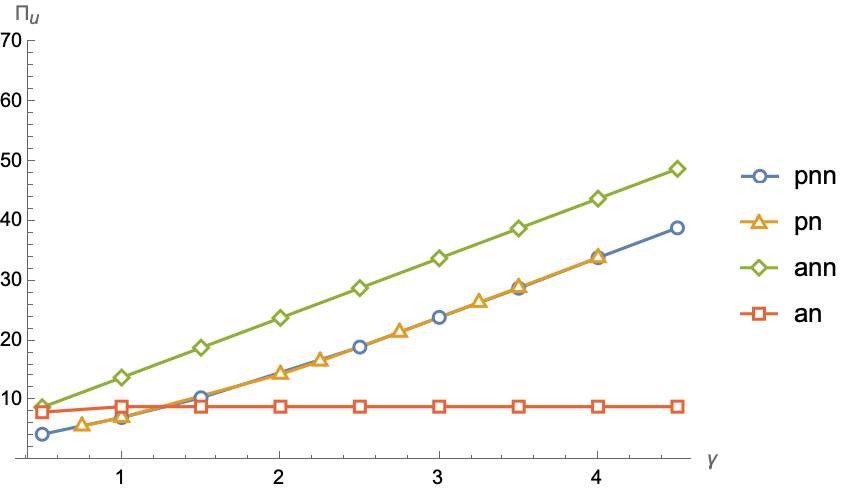}
\caption{Access ISP's profit as a function of $\gamma$}\label{fig:profitispgammaplot}
\end{center}
\end{figure}

\begin{figure}[thb]
\begin{center}
\includegraphics[width=0.6\textwidth]{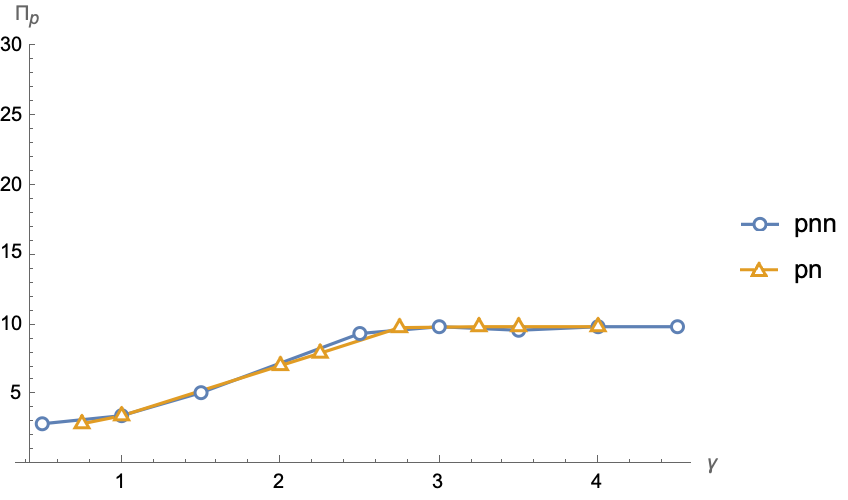}
\caption{Platform's profit as a function of $\gamma$}\label{fig:profitplatformgammaplot}
\end{center}
\end{figure}

\begin{figure}[t]
\begin{center}
\includegraphics[width=0.6\textwidth]{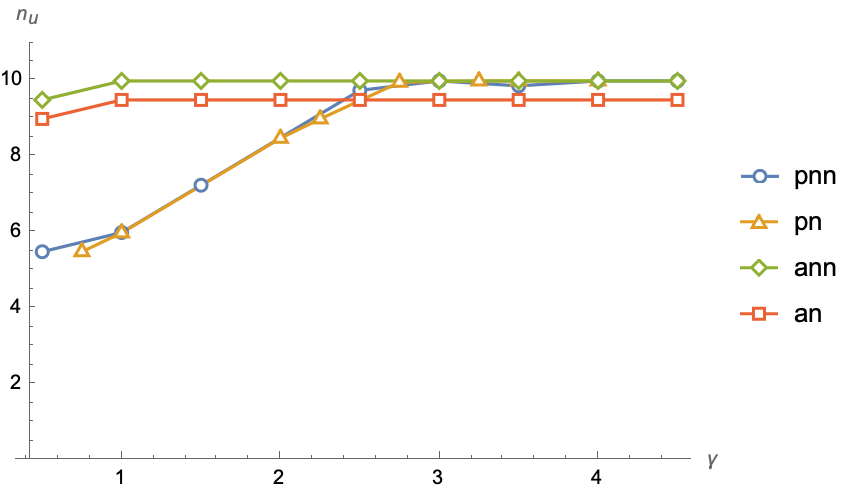}
\caption{Number of subscribers as a function of $\gamma$}\label{fig:nugammaplot}
\end{center}
\end{figure}

\begin{figure}[t]
\begin{center}
\includegraphics[width=0.6\textwidth]{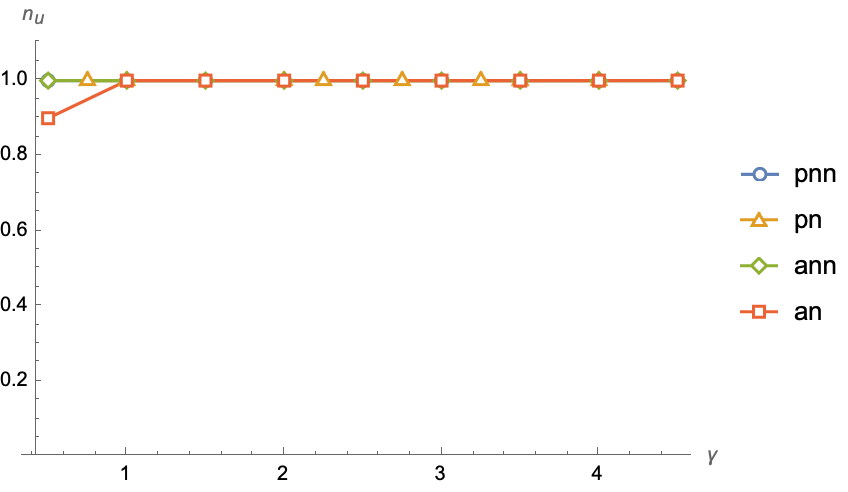}
\caption{Number of joined CPs as a function of $\gamma$}\label{fig:ncgammaplot}
\end{center}
\end{figure}

\begin{figure}[t]
\begin{center}
\includegraphics[width=0.6\textwidth]{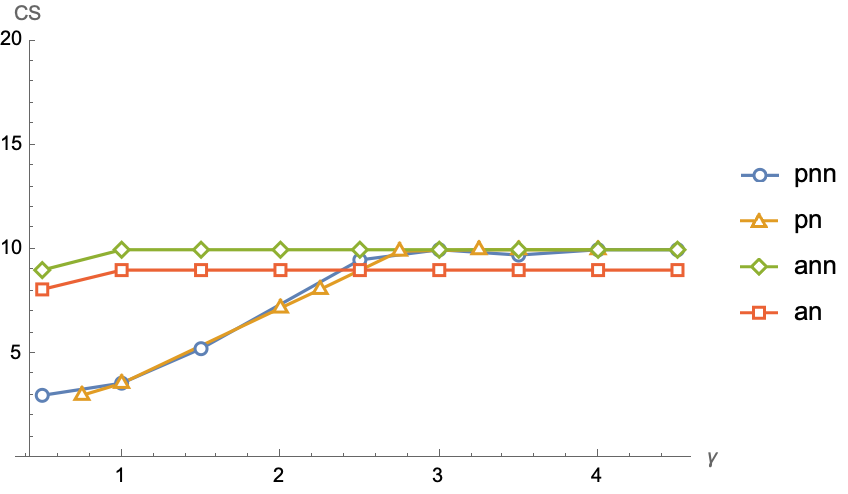}
\caption{Consumers' surplus as a function of $\gamma$}\label{fig:CSgammaplot}
\end{center}
\end{figure}

\begin{figure}[t]
\begin{center}
\includegraphics[width=0.6\textwidth]{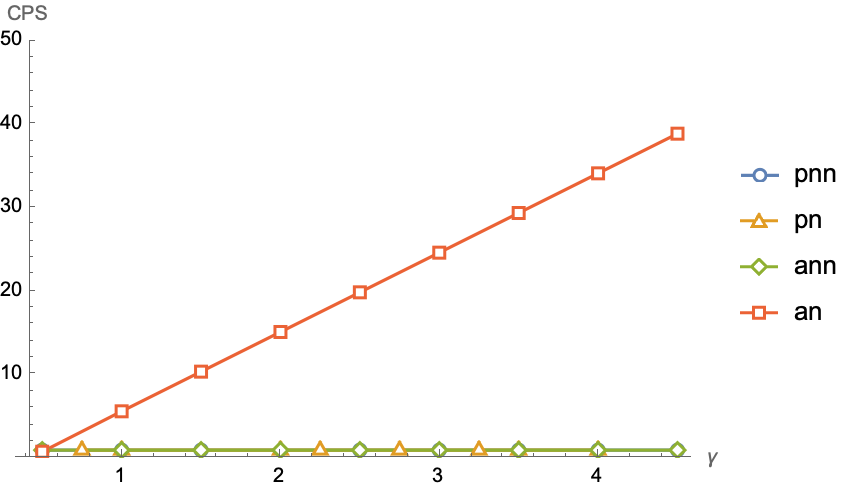}
\caption{CPs surplus as a function of $\gamma$}\label{fig:CPSgammaplot}
\end{center}
\end{figure}

\begin{figure}[t]
\begin{center}
\includegraphics[width=0.6\textwidth]{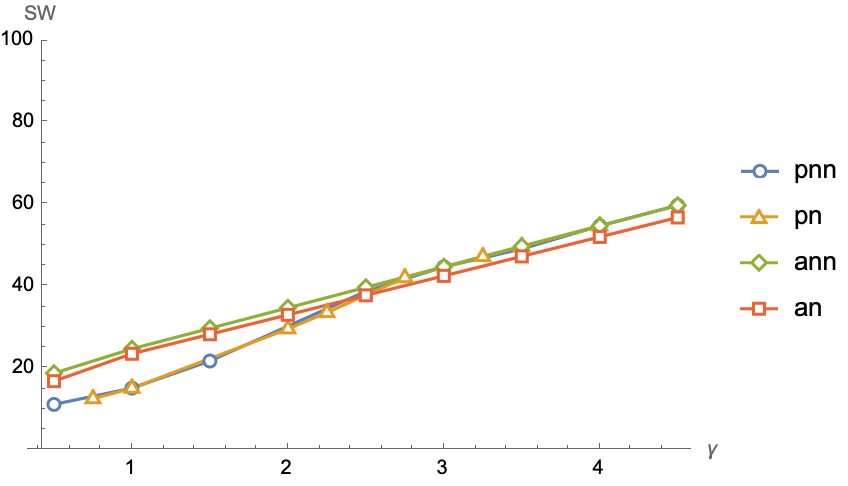}
\caption{Social welfare as a function of $\gamma$}\label{fig:SWgammaplot}
\end{center}
\end{figure}

We discuss the results in terms of ISP profits (Fig.~\ref{fig:profitispdeltaplot} and Fig.~\ref{fig:profitispgammaplot}), platform profits (Fig.~\ref{fig:profitplatformdeltaplot} and Fig.~\ref{fig:profitplatformgammaplot}), number of subscribers (Fig.~\ref{fig:nudeltaplot} and Fig.~\ref{fig:nugammaplot}), number of joined CPs (Fig.~\ref{fig:ncdeltaplot} and Fig.~\ref{fig:ncgammaplot}), users/consumers' surplus (Fig.~\ref{fig:CSdeltaplot} and Fig.~\ref{fig:CSgammaplot}), CP surplus (Fig.~\ref{fig:CPSdeltaplot} and Fig.~\ref{fig:CPSgammaplot}) and social welfare (Fig.~\ref{fig:SWdeltaplot} and Fig.~\ref{fig:SWgammaplot}).

The Consumers' and CPs' surpluses are respectively computed as follows:
\begin{align}
CS & \equiv N_u \int_0^u u \, 1 \, du_0\\
CPS & \equiv N_c \int_0^{\Pi_c} \Pi_c \,  1 \, d\Pi_0.
\end{align}
An the Social Welfare is the sum of the surpluses of all agents:
\begin{equation}
SW = CS + CPS + \Pi_p + \Pi_u.
\end{equation}

The parameters used are $N_u=10$, $N_c=1$, $r_u=0.9$, $r_c=0.9$, $\delta=2$, $\gamma=4$ and $a=0.5$ if not stated otherwise.  

We conduct below comparative statics, that is, we characterize the equilibrium of the three-stage game described above as one parameter is varied across a range of values. Specifically, we analyze the effect of parameter $\delta$, which characterizes how intense the CP cross-network effect is; and of parameter $\gamma$, which quantifies the per subscriber advertising revenue for the CP.

We analyze the results comparatively between four possible scenarios according whether the access ISP is subject to net neutrality regulation or not; and whether the platform is present or absent:
\begin{itemize}
\item the platform is present and the ISP is non-neutral  (\emph{pnn} scenario);
\item the platform is present and the ISP is neutral (\emph{pn} scenario);
\item the platform is absent and the ISP is non-neutral (\emph{ann} scenario);
\item the platform is absent and the ISP is neutral (\emph{an} scenario)
\end{itemize}

The results will show that, for an intermediate range of $\delta$ and $\gamma$ values, the order of preference of the scenarios for each agent is the one shown in Table~\ref{tab:summary}.

\begin{table}[bt]
\begin{center}
\begin{tabular}{r|c|c|c|c}
    & pnn & pn & ann & an \\
 \hline 
 Access ISP   & 2nd-3rd & 2nd-3rd & 1st & 4th  \\
 Platform   & 1st-2nd & 1st-2nd & 3rd-4th & 3rd-4th  \\
 Users   & 1st-4th & 1st-4th & 1st-4th & 1st-4th  \\
 CPs   & 2nd-4th & 2nd-4th & 2nd-4th & 1st  \\
 Social welfare    & 1st-3rd & 1st-3rd & 1st-3rd & 4th 
\end{tabular}
\caption{Scenario comparison}\label{tab:summary}
\end{center}
\end{table}

These results can be summarized as follows:
\begin{itemize}
\item The users are indifferent between the four scenarios. And the CPs strictly prefer the scenario where there is no platform and a neutral ISP is operating, since this means not paying to neither the platform nor the access ISP. And finally, the social welfare is the same and greater in all scenarios but the one where the platform is absent and the ISP is neutral.
\item When a non-neutral ISP is operating, the presence of the platform is not relevant for the CPs; however, when a neutral ISP is operating, the presence of the platform worsens the situation for the CPs. One can conjecture that the strategical interaction between the platform and the access ISP is more beneficial for the CPs when the play field is level (i.e., both platform and access ISP can adjust the same number of strategies, one price at each side). 
\item Trivially, the access ISP prefers to apply two-sided pricing and that the platform is absent.
\item Finally, the platform prefers to be present than to be absent, but it is indifferent between interacting with a non-neutral and with a neutral ISP. This means that the platform pricing is so flexible that it can capture the same value against a non-neutral ISP as against a neutral ISP.
\end{itemize}

\subsection{Comparative statics: \texorpdfstring{$\delta$}{delta}}

The parameter $\delta$ varies between $1$ and $2.5$. 

The results confirm that, for all $\delta$ values, the order of preference of the scenarios is the one shown in Table~\ref{tab:summary}.

\begin{figure}[t]
\begin{center}
\includegraphics[width=0.6\textwidth]{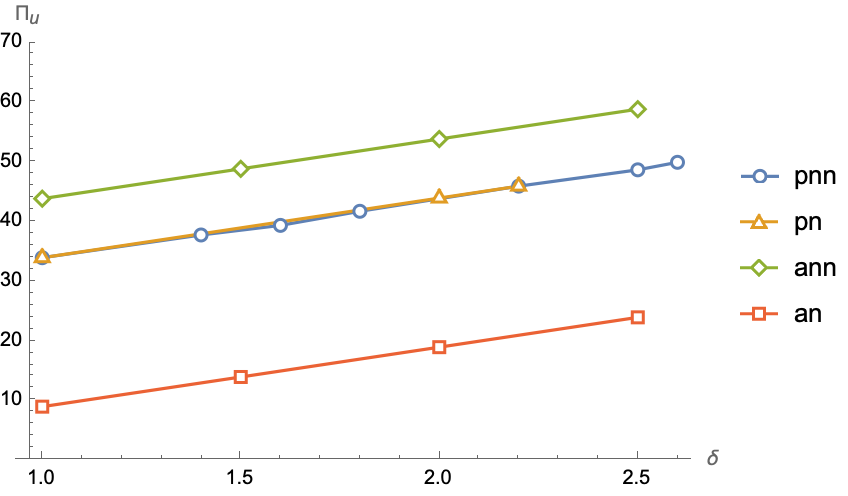}
\caption{Access ISP's profit as a function of $\delta$}\label{fig:profitispdeltaplot}
\end{center}
\end{figure}

\begin{figure}[t]
\begin{center}
\includegraphics[width=0.6\textwidth]{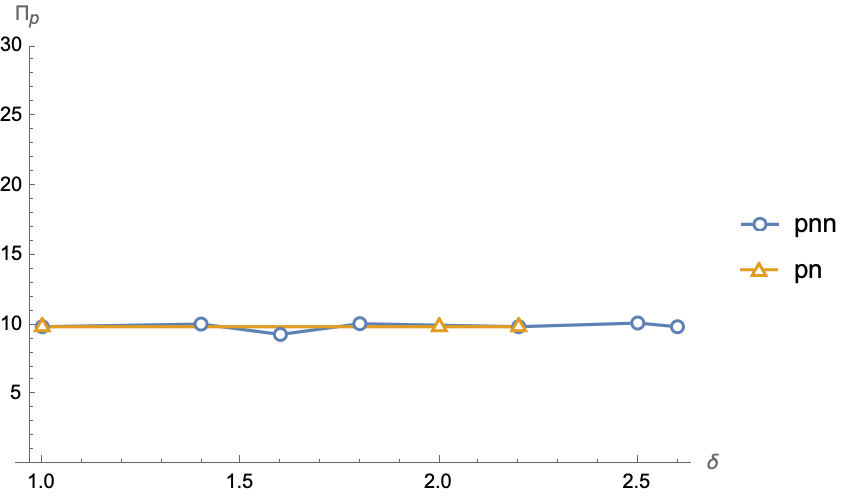}
\caption{Platform's profit as a function of $\delta$}\label{fig:profitplatformdeltaplot}
\end{center}
\end{figure}

Additionally, as $\delta$ increases, which means that the network effect that the CPs exert on the user side strengthens, the access ISP is able to capture the value associated with this increase (Fig.~\ref{fig:profitispdeltaplot}), but the platform cannot (Fig.~\ref{fig:profitplatformdeltaplot}), and therefore high values of $\delta$ are the preferred ones by the access ISP.


\FloatBarrier

\subsection{Comparative statics: \texorpdfstring{$\gamma$}{gamma}}

The parameter $\gamma$ varies between $0.5$ and $4.5$. .

The results confirm that, for $\gamma$ values larger than 3, the order of preference of the scenarios is the one shown in Table~\ref{tab:summary}.

Additionally, as $\gamma$ is forced to have low values, which means that the revenue per user that each CP gets is reduced, the users cannot be retained at the full participation level, and this is specially so for the scenarios with platform present, which switch from 1st-3rd option to the 3rd-4th option as far as $\Pi_u$, $CS$ and $SW$ are concerned.

\section{Conclusions and limitations}\label{sec:conclusions}

We conclude preliminary that:
\begin{itemize}
\item the presence of a platform does not modify the incentives of the CPs when the ISP is not subject to net neutrality regulations. The social welfare is unaffected by the platform presence.
\item but, when the ISP is subject to such regulation, the presence of a platform makes the CPs worse off, and the social welfare is also unaffected by the platform presence.
\end{itemize}

These conclusions reveal the fact that the platforms are extracting surplus from the CPs under the current net neutrality regime for the ISP, and that the platforms would not be able to do so under the counter-factual situation where the ISPs could apply two-sided prices.

We acknowledge the following limitations of the conducted research:
\begin{itemize}
\item The decision of modeling the CPs as non-atomic agents limits the applicability of our study to the settings where the CPs are small compared with the platform and the ISPs. Scenarios where the platform not only intermediates but also provides contents, such as YouTube, which typically cannot be assumed to be non-atomic, are not appropriately matched by our model. The analysis of the latter would need a different model for the CPs: a small number of CPs would be modeled (one or two) and some or all CPs would be integrated with the platform.
\item Partly as a consequence of the above limitations, the sort of net neutrality regulation that is explored in this work is restricted to what is commonly referred to  ``strong neutrality''~\citep{maille2022}, i.e., the very same fact that a fee is charged by the access ISP to the CPs, and leaves out some other types of neutrality regulation. For instance, under the name ``weak neutrality'', a side payment would be allowed provided that it is charged to all CPs under nondiscriminatory terms. The analysis of the latter scenario would need the modeling of the CPs as atomic players in the game model.
\item Finally, we have assumed that the CP outside option does not allow reaching the users, which is very restrictive under the current multiplicity of interconnection alternatives offered by big CPs, Tier-1 ISPs and Cloud Providers~\citep{stocker2017}. The inclusion of these richer set of alternatives is under consideration for further study. 
\end{itemize}
 
\bibliography{bib}

\end{document}